\documentclass[aps,twocolumn,amsmath]{revtex4}
\newcommand{\be}{\begin{eqnarray}}
\newcommand{\ee}{\end{eqnarray}}

\newcommand{\lb}{\label}
\def\>{\rangle}
\def\<{\langle}

\def\li{{\mathcal L}}
\def\pp#1{\partial_{#1}}
\def\lb{\label}
\begin{document}
\title{Exact Black Hole Solutions in Noncommutative Gravity}
\author{
Peter Schupp and Sergey N. Solodukhin$^*$
} \affiliation{{\it Jacobs University Bremen, Campusring 1, Bremen
28759, Germany\\
${}^*$Universit\'e de Tours, Parc de Grandmont, 37200 Tours,
France}}

\begin{abstract}
\noindent {An exact spherically symmetric black hole solution
of a recently proposed noncommutative gravity theory based on star
products and twists is constructed. This is the first nontrivial
exact solution of that theory. The resulting noncommutative black
hole quite naturally exhibits holographic behavior; outside
the horizon it has a fuzzy shell-like structure, inside the horizon
it has a noncommutative de Sitter geometry. The star product and
twist contain Killing vectors and act non-trivially on tensors except the
metric, which is central in the algebra.
The method used
can be applied whenever there are enough spacetime symmetries.
This includes noncommutative versions of
rotating and charged black holes and  higher-dimensional and cosmological solutions.\\[1ex]
}
\end{abstract}
\vskip 2.pc \maketitle

{\it Introduction.}---It is generally expected that the interplay of
quantum and gravitational effects at small distance scales can lead
to a quantization of spacetime itself: Einstein's theory of gravity,
general relativity, should be replaced by a new theory in which the
metric, coordinates and other building blocks of classical geometry
are promoted to operator valued objects. Such a generalized gravity
theory does not necessarily have to strive to be a candidate for a
fundamental theory of physics; important physical insights can also
be obtained from a phenomenological description that is
nevertheless capable of capturing essential aspects of the quantum
geometry that is expected to play a role in any reasonable theory of
quantum gravity~\cite{DeWitt:2007mi}. An ideal theoretical laboratory to test such ideas
are quantum black holes~\cite{thooft,susskind}. Motivated by this,
a lot of effort has gone into the study of noncommutative
generalizations of quantum field theory. Surprisingly, when applied
to gravity itself this has not been very successful so far even
though gravity was the original motivation for the endeavor: It
turns out to be very difficult to consistently construct a
noncommutative theory of gravity.
The main reason is the peculiar nature of the group of
diffeomorphisms that does not allow a simple noncommutative
extension. This is in contrast with the group of internal gauge
symmetries: Mathematically consistent theories of noncommutative
gauge fields are readily available and emerge in fact quite
naturally in string theory.

Only recently, it has been attempted to formulate theories of
noncommutative gravity that preserve symmetries
of classical general relativity.
In~\cite{Aschieri:2005yw} a consistent theory of noncommutative gravity based
on star products and a twist construction
that allows to retain the full diffeomorphism invariance of
general relativity was proposed.
Among other approaches let us mention the closely related
work~\cite{Chamseddine:2000si} that features a noncommutative complex metric.
Other authors have concentrated on noncommutative effects coming from gravity
sources. For reviews see~\cite{Szabo:2006wx,Nicolini:2008aj}. Noncommutative spacetimes 
consistent with \cite{Aschieri:2005yw} have been studied in \cite{Ohl:2008tw}.

Despite the fact that theories of noncommutative gravity are only
starting to mature, it has became  quite popular in the literature
to discuss ``noncommutative black holes" or ``noncommutative
horizons" (and numerous implications of noncommutativity for black
hole physics) based mostly on heuristic arguments. Frequently these
objects do not have a clear mathematical status and moreover it is
not clear whether and how they can emerge as solutions to a theory
of gravity dealing with coordinates and metric as noncommutative
operators.

In this letter we fill a gap in this discussion and analyze exact
solutions to the theory of noncommutative gravity proposed
in~\cite{Aschieri:2005yw}. We analyze in particular a class of such solutions with
rotational symmetry. This restriction provides already in
the commutative case an important class of spacetimes including
black holes. We  outline how to construct more general
noncommutative black hole and cosmological solutions.

A key point in our discussion is the observation that the
noncommutative structure of spacetime given  generically by the
relation
\begin{equation}
[\hat x^\mu, \hat x^\nu]= i \theta^{\mu\nu} (x) \;,\lb{1}
\end{equation}
should be consistent with the symmetry of spacetime just like the
metric tensor is. Thus, in the case of the spherical symmetry not only the metric
$g_{\mu\nu}(x)$ but also the tensor $\theta^{\mu\nu}(x)$ should obey spherical symmetry.
Choosing an appropriate coordinate system
$(x_0,x_i,\ i=1,2,3)$ we then have the general form for the
commutation relation
\begin{equation}
[\hat x_i, \hat x_j]= i \lambda(\hat x^2)  \epsilon_{ijk} \hat x_k, \lb{2}
\end{equation}
where $\lambda$ can be a function of the Casimir $\hat x^2=\sum_i \hat x_i \hat x_i$
of the group of rotations. There is an element of incompleteness in
this picture. The way we think about noncommutative gravity
suggests that both $\theta^{\mu\nu}$ and $g^{\mu\nu}$ are to be
considered as independent dynamical variable. However, although the
dynamic of the metric is supposed to be determined from a
noncommutative version of Einstein's equations, no such equations
are known for the parameter of noncommutativity $\theta^{\mu\nu}$.
Some sort of equations, analogous to those for the metric,  are
however anticipated so that the matter distribution would determine
not only how spacetime is curved but also how it deviates from
the classical commutative structure.  Such a complete theory is not
available at the moment. Our approach, relying on the symmetry of
the problem, nevertheless allows to maximally restrict the structure
of the exact solution by the given set of equations. That not everything is
fixed by  symmetry is however manifest in the fact that the form of
the function $\lambda(x^2)$ is not completely determined -- we
necessarily have an ambiguity in our solutions. Additional ideas (like holography)
can be used to fix this ambiguity. This does
not preclude of course the necessity of eventually finding the
complete set of equations.

\noindent {\it Deformed gravity with twisted symmetry.}
Classical diffeomorphisms are not compatible with a
noncommutative product of tensors. Not even the product of
two scalars will in general transform
as a scalar. A deformed notion of symmetry is needed.
It was shown in \cite{Aschieri:2005yw} that one can in fact keep the
classical symmetry algebra---the universal enveloping algebra of the
vector fields---and deform the Leibnitz rule, i.e.\ the coproduct.
Such a deformation is most conveniently done by a Drinfel'd twist.
The concrete realization presented in
\cite{Aschieri:2005yw} uses an abelian Reshetikhin-type twist
and works for commutation relations in which $\theta^{\mu\nu}$ is constant.
This framework is too restrictive for our present purposes, we therefore
suggest a generalization of this structure which is based on a preferred
 set of vector fields, $\xi_n$. The suggested form
of the twist in our construction is an expansion in powers of the Lie derivatives with respect
to the Killing vectors acting on a product of two Hilbert spaces:
\begin{equation}
{\cal F}=\sum_{N,M} C_{n_1..n_N m_1..m_M} {\cal
L}_{n_1}..{\cal L}_{n_N}\otimes {\cal L}_{m_1}..{\cal L}_{m_M}
\lb{twist}
\end{equation}
where ${\cal L}_{m}$ is the Lie derivative
with respect to the Killing vector $\xi_m$. The coefficients
$C_{n_1..n_N m_1..m_M}$ do not need to be constants as
long as they commute with the vector fields $\xi_m$.
A generalized star product is obtained by letting the twist act
on a pair of tensors before multiplying the resulting expression
with the usual point-wise product. The twist and the
corresponding star product will in general act non-trivially on
tensors. The generalized star product encodes all the information of a
noncommutative (differential) tensor calculus. The resulting
twisted tensor calculus is governed by
a few simple rules: The transformation of individual tensors
is undeformed; tensors and other geometric objects must be star-multiplied;
products of tensors are transformed using the twisted Leibnitz rule.
The covariant derivative becomes
\begin{equation}
D_\mu\star V^\nu = \partial_\mu V^\nu +
V^\alpha \star \Gamma_{\mu\alpha}^\nu \;.
\end{equation}
Placing the connection coefficients to the right of the contravariant vector in this
expression represents a choice. Placing the coefficients to the left also yields a
covariant expression and is in fact the natural choice for covariant vectors.
Here and also in some of the following expressions, in particular for the metric,
we differ from the original article~\cite{Aschieri:2005yw} and some of the subsequent
publications based on it, where covariance was not
studied carefully enough for our present purposes.

The noncommutative expressions for curvature and torsion are
\begin{equation}
[D_\mu\stackrel{\star}{,}D_\nu] \star V^\rho = V^\sigma \star R^\rho{}_{\sigma\mu\nu}   -
T_{\mu\nu}^\lambda \star (D_\lambda \star V^\rho) \;.
\end{equation}
In components:
\begin{align}
R^\rho{}_{\sigma\mu\nu} & =
\partial_\mu \Gamma_{\nu \sigma}^\rho -
\partial_\nu \Gamma_{\mu \sigma}^\rho +
\Gamma_{\nu\sigma}^\lambda \star \Gamma_{\mu\lambda}^\rho -
\Gamma_{\mu\sigma}^\lambda \star \Gamma_{\nu\lambda}^\rho \\
T_{\mu\nu}^\lambda & = \Gamma_{\mu\nu}^\lambda -
\Gamma_{\nu\mu}^\lambda
\end{align}
{\it Remark:} These expressions generalize the results of~\cite{Aschieri:2005yw} to the case
of non-constant $\theta$, and implicitly use the fact that
$\star \pp\mu = \pp\mu = \pp\mu\star$ i.e $\star d = d = d \star$ and therefore $d\star d = d^2 = 0$,
which all follows from
$[\li_V, d] = 0$. Also note that all ordering ambiguities that require us to make choices
disappear in the solutions that we shall discuss.

Metric compatibility of the connection and the two conditions
\begin{equation}
\left(G_{\mu\nu}\right)^* = G_{\nu\mu}, \qquad \left(\Gamma_{\mu\nu}^\lambda\right)^* = \Gamma_{\nu\mu}^\lambda \;,
\end{equation}
which hold in particular for a real symmetric metric, fix the Christoffel
connection coefficients in terms of the metric
\begin{equation}
\Gamma_{\alpha\beta\gamma} =  \frac{1}{2}\left(\partial_\alpha
G_{\gamma\beta} +
\partial_\beta G_{\alpha\gamma} -
\partial_\gamma G_{\alpha\beta} \right)\;,
\end{equation}
where the last index can be raised by solving
(in a formal series in $G_{\alpha \beta}$ and using vanishing torsion) the following
expression:
\begin{equation}\label{christoffel}
\Gamma_{\alpha\beta\gamma} = \frac{1}{2}\left(G_{\gamma\lambda} \star \Gamma_{\alpha\beta}^\lambda
+ \Gamma_{\beta\alpha}^\lambda \star G_{\lambda\gamma}\right) \;.
\end{equation}

In terms of the hermitean (or real symmetric) Christoffel connection we can write a noncommutative analog of the Geodesic equation
\begin{equation}\label{geodesic}
    \frac{d u^\gamma}{d\lambda} = u^\alpha \star \Gamma_{\alpha\beta}^\gamma \star u^\beta\;, \qquad u^\alpha = \frac{dx^\alpha}{d\lambda}\;,
\end{equation}
which should be interpreted as a quantum mechanical Heisenberg-type evolution equation for the operators $u^\alpha$, $x^\alpha$ that describe the possible location of events. Alternatively and in fact more generally one can of course also study fields in the noncommutative background.

Contracting two indices of the Riemann curvature tensor in the usual way gives the covariant
Ricci tensor
\begin{equation}
R_{\mu\nu} = R^\lambda{}_{\mu\lambda\nu}\;.
\end{equation}
The noncommutative gravity equations in vacuum become
\begin{equation}
R_{\mu\nu}
= 0 \;.
\end{equation}
In the presence of matter the noncommutative gravity equations are best written
in the form that Einstein originally used,
\begin{equation}\label{ncgravitymatter}
R_{\mu\nu} = 8 \pi G_N \left(T_{\mu\nu} - \frac{1}{2} G_{\alpha\beta} T\right) \;,
\end{equation}
where $T_{\mu\nu}$ and its trace $T$ describe (noncommutative) matter.
Expanded in terms of the metric field and its derivatives these are very
complicated PDEs with derivatives up to arbitrary order.
A priory the question for solutions and appropriate boundary conditions of these PDEs seems to be ill-posed.
We define a solution to be a mutually compatible pair of an algebra (with compatible twist)
and a metric that solve the non-commutative gravity equations.

{\it Noncommutative Schwarzschild solution.}
As we are interested in spherical symmetry, we consider a set of Killing vectors
$\xi_i=\epsilon_{ijk}x^k\partial_j$ that satisfy
\begin{equation}
[\xi_i,\xi_j]= -\epsilon_{ijk}\xi_k \lb{Killing} \qquad {\mathcal L}_{\xi_i} g^{\mu\nu} = 0
\end{equation}
(plus one Killing vector for time) and a compatible algebra
\begin{equation}\label{algebra}
    [x_i \stackrel{\star}{,} x_j] = 2 i \lambda \epsilon_{ijk} x_k
\end{equation}
where $\lambda$ can be a function of the Casimir, as we have mentioned already.
For the time being we shall concentrate on the case of commutative time in the spirit of
looking for stationary solutions.
Later in the paper we shall also discuss the possibility of time-space noncommutativity in our model.
For consistency we need to use isotropic coordinates in the ansatz for the metric:
\begin{equation}\label{metricansatz}
  ds^2 = -A(\rho) dt^2 + B(\rho)(dx^2 + dy^2 + dz^2) 
\end{equation}
where $\rho^2 \equiv g_{ij} x^i x^j = x^2 + y^2 + z^2$ is the square of the radial coordinate on the classical commutative auxiliary space that underlies the noncommutative star product algebra and $A(\rho)$, $B(\rho)$ are functions to be determined.

The noncommutative coordinate algebra is realized (for $x\neq 0$) by the star product~\cite{Presnajder:1999ky}
\begin{multline}\label{grossestar}
f \star g = f g  +
\sum_{n=1}^\infty C_n({\lambda \over \rho}) J^{a_1
b_1}\ldots J^{a_n b_n} \cdot \\ \cdot \partial_{a_1} \ldots \partial_{a_n}f \,
\partial_{b_1} \ldots \partial_{b_n}g \;,
\end{multline}
where
\begin{equation}
J^{ab} = \rho^2 \delta_{ab} - x_a x_b + i \rho \epsilon_{abc}x_c
\end{equation}
and the coefficients $C_n$ are related to the beta function:
\begin{equation}
\begin{split}
C_n({\lambda \over \rho}) & =  B(n,{\rho \over \lambda}) \\
& =  \frac{\lambda^n}{n! \,{\rho} ({\rho} - \lambda) ({\rho} -
2\lambda)\cdots ({\rho}-(n-1)\lambda)}  \;.
\end{split}
\end{equation}
An alternative form of the same star product is~\cite{Alekseev}
\begin{equation}
f \star g = f g + \sum_{n=1}^\infty C_n({\lambda \over \rho})
{\xi_+}^n f \, {\xi_-}^n g
\end{equation}
with left invariant Killing vector fields $\xi_{\pm}$
that are fixed by the relation $\xi_{\pm} \equiv \xi_1 \pm i \xi_2$ which holds
along the positive $z$-axis.
In this form the star product for functions can be generalized
to a star product of tensors with the help of Lie derivatives
\begin{equation}\label{startensor}
    V \star W = V W + \sum_{n=1}^\infty C_n({\lambda \over \rho})
{\mathcal L}_{\xi_+}^n V \, {\mathcal L}_{\xi_-}^n W \;.
\end{equation}
A corresponding projective twist of the anticipated form can be read off: 
\begin{equation}\label{twist2}
    \bar{\mathcal F} = \sum C_n(\frac{\lambda}{\rho}) {\mathcal L}_{\xi_+}^n
\otimes  {\mathcal L}_{\xi_-}^n\;.
\end{equation}
This twist has previously been discussed in \cite{Kurkcuoglu:2006iw}, where it was called a ``pseudo twist'' because it defines a partial isometry rather than a bijection. This property will be responsible for the projection onto an onion-type spacetime which effectively reduces the dimensionality of spacetime by one as we shall see.

Notice that the components of the Killing vectors are linear in the
coordinates. An immediate consequence of this property is the fact
that any number of partial derivatives
$\partial_{\sigma_1}..\partial_{\sigma_p} T_{\mu_1..\mu_k}$ applied
to a tensor $T_{\mu_1..\mu_k}$ transforms as a tensor under the
infinitesimal transformations generated by vectors $\xi_i$. A
related fact that is checked by direct calculation is that the Lie
derivative ${\cal L}_i$ w.r.t. vector $\xi_i$ commutes with any
partial derivative,
\begin{equation}
{\cal L}_i\partial_\sigma
T_{\mu_1..\mu_k}=\partial_\sigma{\cal L}_iT_{\mu_1..\mu_k} \lb{comm} \;,
\end{equation}
which in fact is just a coordinate dependent expression of the
relation $[{\cal L}_i , d] = 0$.
The spherically symmetric metric is by definition annihilated by
each Killing vector, ${\cal L}_i g_{\mu\nu}=0,\ i=1,2,3$. The
property (\ref{comm}) means that applying any number of partial
derivatives to metric we get an object that is annihilated by the
Killing vectors,
\begin{equation} {\cal L}_i
\partial_{\sigma_1}..\partial_{\sigma_p} g_{\mu\nu}=0
\lb{ptg}
\end{equation}
In particular, this is true for the
Christoffel symbols and the Riemann tensor. Now, since our star
product is defined with help of the twist (\ref{twist},\ref{twist2}) is built in
terms of the Lie derivatives of the Killing vectors, we have the
following property
\begin{equation}\lb{*}
\begin{split}
T_{\mu_1..\mu_k}\star\partial_{\sigma_1}..\partial_{\sigma_p}
g_{\mu\nu} & =T_{\mu_1..\mu_k}\partial_{\sigma_1}..\partial_{\sigma_p}
g_{\mu\nu}\\
& = \partial_{\sigma_1}..\partial_{\sigma_p} g_{\mu\nu}\star T_{\mu_1..\mu_k}
\end{split}
\end{equation}
valid for any tensor
$T_{\mu_1..\mu_k}$. Thus, the star product  disappears  and the
gravitational equations without matter become commutative, i.e. those derived by
Albert Einstein in 1915. The spherically symmetric solution of these
equations was derived by Schwarzschild in 1916. We obtain it in isotropic coordinates
in the following form
\begin{equation}
g_{ 00}={a\over r}-1 \ , \ \ g_{0i}=0 \ , \ \ g_{ij}={r^2\over
\rho^2} \delta_{ij} \lb{solution}
\end{equation}
where
\begin{equation}
 r={1\over \rho}\left(\rho+{a\over 4}\right)^2~.
\end{equation}
The parameter $a$ is the Schwarzschild radius, $r$ is the Schwarzschild radial coordinate.
The event horizon is at $r=a$ or, in our coordinates, at
$\rho={a\over 4}$. The important difference of metric
(\ref{solution}) as a solution to the noncommutative gravitational
equations is that now $\rho$ is {\it not} a continuous number but a function of
the quadratic Casimir operator of the noncommutative algebra of coordinates. In fact,
using (\ref{grossestar}) we find
\begin{equation}\label{rhocasimir}
    \sum x_i\star x_i = \rho(\rho+2\lambda)\;.
\end{equation}
Spacetime ``coordinates'' and fields (other than the metric) are nontrivial operators acting on a Hilbert space. Measuring the coordinates of an event should give real results, hence we should consider unitary representations of the coordinate algebra (\ref{algebra}). These are labeled in terms of spin $j=0,1/2,1,3/2,\ldots$ and the corresponding eigenvalues of the Casimir $\sum x_i\star x_i$ are $(2\lambda)^2 j(j+1)$. In terms of the radial coordinate $\rho$ this becomes
\begin{equation}\label{quantizedrho}
    \rho = n \lambda \,, \qquad n \equiv 2 j = 0,1,2,\ldots
\end{equation}
 with standard degeneracy $(2j+1) = n + 1$. The condition $\rho > a/4$ ensures that we are outside the horizon and puts a lower limit on the quantum number $n$: $n > a/4\lambda$. This completes our derivation of the exact solution.

The noncommutative spacetime described by the metric
(\ref{solution}) is sliced with fuzzy spheres. The radial coordinate has become discrete, effectively
decreasing the dimensionality of spacetime by one. For constant $\lambda$ we would have an equidistant
radial spectrum (in isotropic coordinates), but $\lambda$ can itself  be a function of $\rho$ and conceivably
also of the black hole parameter $a = 2m$. The possible form of the function $\lambda
(\rho)$ can be restricted by physical considerations: A natural choice leads to the equidistant area spectrum conjectured by Bekenstein and Mukhanov \cite{Bekenstein:1995ju}. For large distances $\rho/a \gg 1$ there are choices for $\lambda$ that lead to a constant density of states. A more detailed discussion of the physics and mathematics underlying $\lambda$ will be given elsewhere.

It is instructive to look in more detail
at the Hilbert space of states that describes the possible location of events in the noncommutative black hole spacetime: Without black hole, i.e. for $a=0$ all values of the quantum number $n=2j$ are allowed, each value corresponding to a unitary irreducible representation of SU(2) of dimension $2j+1$.
For nonzero $a$ all states with $n \leq a/4 \lambda$ are effectively hidden by the horizon. This is not meant to imply that these states continue to exist inside the horizon -- in fact the structure inside the horizon is quite different, as outlined below -- but are in any case not accessible for outside observers (resting sufficiently far away from the black hole). The Hilbert space of the missing states can be rewritten as a direct sum of two representations: One is the tensor product of two irreps of spin $j$ and correspond  to a scalar function on a fuzzy sphere; the other one is the tensor
product of a spin $j$ and a spin $j\pm 1/2$, corresponding to a spinor on a fuzzy sphere. The whole structure is reminiscent of a fuzzy supersphere \cite{GrosseReiter}.
The Hilbert space of the accessible states outside the horizon is thus given by a scalar plus a spinorial function on a sphere minus a fuzzy sphere. We are quite literally dealing with a fuzzy black hole. The horizon  is modeled by a fuzzy sphere -- the fact so much
anticipated in the literature. It is the fuzzy sphere of states that ceases to exist due to the formation of the black hole. 
The surprising thing is that the probability amplitude for the location of events is described by (wave)functions on the two dimensional surface of a sphere, while in ordinary quantum mechanics one would have expected wave functions living in three dimensions. The idea of holography is realized very naturally in our model. A heuristic explanation of the fact that the dimension is effectively reduced from 3+1 to 2+1 is that the coordinate commutation relations (\ref{algebra}) allow to express any one coordinate in terms of the other two. This effect disappears in the commutative limit.


\noindent{\it Inside the horizon.} Finding a solution of the noncommutative gravity
equations inside the horizon is more involved. The purpose of finding a solution is on one hand a proof of principle on the other hand it is interesting to see what happens with the central singularity.
We shall focus again
on the case of commutative time. 
The noncommutative
solution outside the horizon is based on the possibility to choose
appropriate Euclidean coordinates $x^1$, $x^2$ and $x^3$ that
form an SU(2) algebra. This possibility is related to the fact that
outside the horizon, where $g_{00}>0$,  the slices of constant time are
conformal to  Euclidean flat spacetime. Thus, in this case, the
construction  goes in the same way as that of fuzzy spheres in flat
space. Surprisingly, this construction cannot easily be continued
inside the horizon, where $g_{00}<0$, because
there are no slices that are conformal to 3-dimensional Euclidean
flat space. Inside the horizon the slices of constant $t$ are conformal
to de Sitter spacetime. Three-dimensional de Sitter spacetime can be
defined starting with flat 4-dimensional spacetime with the metric
\begin{equation}
ds^2=-dx_0^2+dx^2_1+dx^2_2+dx^2_3 \lb{flat}
\end{equation}
and looking at the subspace satisfying the condition
\begin{equation}
-x_0^2+x_1^2+x^2_2+x^2_3=-1 \;.\lb{constr}
\end{equation}
We can now quantize $x_1$, $x_2$, $x_3$ as before, imposing SU(2) commutation relations (\ref{2}).
In terms of $\rho^2=x_1^2+x^2_2+x^2_3$ we find $x_0^2=1+\rho^2$. The metric inside
horizon has the form
\begin{equation}
g_{00}=1-{a\over r}\ , \ \
g_{0i}=0 \, \ \ g_{ij}={r^2\over \rho^2 }(\delta_{ij}-{x_ix_j\over
\rho^2}) \lb{inside}
\end{equation}
The relation between coordinates
$r$ and $\rho$ is
\begin{equation}
\rho^2={a\over 4 r}{1\over (1-{r\over a})} \lb{rrho}
\end{equation}
Infinite values of $\rho$
correspond either to $r=0$ (singularity) or to $r=a$ (horizon). 
We thus need two infinite sequences of fuzzy spheres to cover the 
interior of the black hole with accumulation points at the horizon and at the central singularity.
The singularity at $r=0$ is resolved in
the sense that it is now replaced with a sequence of fuzzy spheres
of quantized radius. The interior of the black hole is mostly
empty. The majority of states is concentrated either near the singularity
or the horizon. This is different from how the states are
distributed outside the horizon where most states are at large values
of the radius.

{\it Time-space noncommutativity.}
In the spirit of a time-independent solution we have so far only considered
noncommutative structures with commutative time.
Spherical symmetry is also consistent with nontrivial time-space commutation relations:
The appropriate commutation relations
\begin{equation}\label{kappa}
    [\hat x_i,\hat t] = i \lambda'(x) \hat x_i
\end{equation}
define what is usually referred to as ``kappa Minkowsky space''. The
parameter $\lambda'(x)$ can depend on the casimir $x^2$, but should be
time-independent. Associativity is a strong constraint in noncommutative
theories. In the present case it implies that commutation relations
(\ref{2}) and (\ref{kappa}) for non-zero $\lambda,\lambda'$
are only consistent if $\lambda$ is proportional to $\hat x^2$,
but in that case the space outside the black hole would consist only of
a single spherical shell. Since this is clearly unphysical,
either $\lambda$ or $\lambda'$ should be zero.
In some approaches to noncommutative gravity it is advantageous to
have non-trivial commutation relations for all spacetime coordinates~\cite{Buric:2008th},
and despite of what we have just discussed, it is indeed possible to find
time-space commutation relations that are consistent with~(\ref{2}):
\begin{equation}\label{kerr}
    [\hat x_i,\hat t] = i\tilde\lambda(x) \epsilon_{ijk} \hat v_j \hat x_k
\end{equation}
or more concisely  $[\vec x,t] = i\tilde\lambda \vec v \times \vec x$.
Without loss of generality $\vec v$ can be chosen to be a unit vector pointing in the
direction of the zenith. The corresponding azimuthal angle $\phi$ satisfies
(loosely speaking) canonical commutation relations with time $[\phi,t] = i\tilde \lambda$.
Mathematically more precisely this should be written
\begin{equation}\label{angletime}
    [\hat t, e^{i\phi}] = \tilde\lambda e^{i\phi}
\end{equation}
and has a very interesting representation theory that requires a careful physical
interpretation which we will discuss elsewhere.
The commutation relations (\ref{kerr}) are not fully spherically
symmetric as there is  a preferred direction---all we have is cylindrical
symmetry. This is, however, exactly what we need to find solutions for noncommutative
rotating  black holes.  For all mentioned commutation relations
there exist twists that have the property that they act trivially on a metric
with the appropriate symmetry. In fact, the commutation relations (\ref{kappa}), (\ref{kerr})
can be identified in a systematic search for ablian twist compatible
with spherical symmetry~\cite{Ohl:2008tw}.
Repeating the construction that we did for the
Schwarzschild black hole with the noncommutative structure (\ref{kerr})
gives a noncommutative version of the Kerr solution. Thus not surprisingly,
the commutation relations (\ref{angletime}) have also been found in a noncommutative
version of the \mbox{(2+1)-}dimensional BTZ black hole~\cite{Dolan:2006hv}.

{\it  Summary and discussion.}---We have outlined a method for the construction of
exact solutions to the noncommutative gravity theory that was
proposed in~\cite{Aschieri:2005yw}. This is not a full quantum theory of gravity
with a quantized gravitational field, but rather a gravity theory in a given
quantum geometric background. Correspondingly we do not consider Hamiltonian constraints
as in canonical quantum gravity~\cite{DeWitt:2007mi}
but instead solve Einstein equations explicitly.
In view of the fact that the deformed Einstein equations are infinite order
partial differential equations for the metric, it was a priori not clear that
exact solutions are possible at all. As in ordinary general relativity the key point are
spacetime symmetries. For the noncommutative Schwarzschild geometry we require not only
the metric but also the noncommutative algebra to be spherically symmetric.
The appropriately generalized star product acts nontrivially on tensors in
such a way that the metric and its derivatives are central in the algebra
of operators. The metric thus turns out to be formally identical to its classical
counterpart (in isotropic coordinates), but coordinates, equations of motion and
field equations are truly noncommutative.
The resulting noncommutative spacetime has a discrete onion-shell like structure
outside the horizon. Counting degrees of freedom of matter in such a background
we quite naturally find holographic behavior with entropy following an area law.
Inside the horizon we find that the central singularity is not part of
the solution, but it is not smeared out as one may have hoped. Such a smearing effect
may nevertheless hold in noncommutative theories, but would most likely come
from deformed matter equations~\cite{Nicolini:2008aj}.


{\it  Acknowledgements.}---We have benefited greatly from discussions with numerous colleagues.
In particular we gratefully acknowledge helpful discussions with Maja Buric, Brandon Carter,
Brian Dolan, Harald Grosse, Kumar Gupta, Denjoe O'Connor,  John Madore and
Christian S\"amann.
One of the authors (PS) thanks Hernando Quevedo, Chryssomalis Chryssomalakos
and Denjoe O'Connor for hospitality at UNAM and DIAS, where part
of this work was done.

\end{document}